\documentclass[a4paper,11pt]{article}
\pdfoutput=1 

\usepackage{jcappub} 

\usepackage[T1]{fontenc} 
\usepackage{epstopdf}
\usepackage{float}
\usepackage{color}
\title{Non-singular flat universes in braneworld and Loop Quantum Cosmology}

\author[a]{Rikpratik Sengupta,}
\author[b]{B C Paul,}
\author[a]{Mehedi Kalam,}
\author[c]{Prasenjit Paul}
\author[d]{and Arkajit Aich}

\affiliation[a]{Department of Physics, Aliah University, Kolkata 700 160, West Bengal, India}
\affiliation[b]{Department of Physics, North Bengal University,\\ Siliguri 734001, West Bengal, India}
\affiliation[c]{Department of Physics, Government College of Engineering and Ceramic Technology,\\ Kolkata 700 010, West Bengal, India}
\affiliation[d]{Department of Basic Sciences, Atria University,\\ Bengaluru 560024, Karnataka, India.}

\emailAdd{rikpratik.sengupta@gmail.com}
\emailAdd{bcpaul@associates.iucaa.in}
\emailAdd{kalam@associates.iucaa.in}
\emailAdd{Pp.phys@gcect.ac.in}
\emailAdd{Arkajitaich1994@gmail.com}

\abstract{In this paper we take matter source with non-linear Equation of state (EoS) that has produced non-singular Emergent cosmology for spatially flat universe in General Relativity and minimally coupled scalar field with two different potentials that produce an inflationary emergent universe for positive spatial curvature in the relativistic context. We study all these three cases both in the context of Randall-Sundrum braneworld and effective Loop quantum cosmology (LQC) for zero spatial curvature that is observationally favoured and in the absence of any effective cosmological constant term. We solve the modified Friedmann equation in each case to obtain the time evolution of the scale factor and use it to check whether the initial singularity can be averted. In almost all the cases we find the initial singularity is absent. We study the nature of the slow roll inflation in the cases where we obtain inflationary emergent universes. The inflationary scenario is found to be improved than in a standard relatvistic context and we compare the improved scenario for both the braneworld and LQC models. Interestingly, we also obtain bouncing and cyclic universes from our analysis in some cases. We find that the initial singularity can be averted for a spatially flat universe with specific choice of matter EoS or scalar field potential, which do not violate the Null Energy condition in most cases, taking into account effective high energy (curvature) corrections with or without extra dimensions.}

\begin{document}
\maketitle
\flushbottom
\section{Introduction} \label{sec1}

Most cosmologists today consider the 'initial singularity'\cite{Hawking} as a major setback of the standard big bang cosmological model. It is geodesically incomplete in the past\cite{Borde}. This might be a limitation of General Relativity (GR) to explain situations involving extremely high spacetime curvature and energy densities as applicable to the primordial universe. There have been a flourish of non-singular cosmological models in the last two decades in the framework of both standard GR as well as its modifications. Non-singular cosmological models can broadly be classified into three categories: (i) the Emergent Universe (EU) scenario\cite{EM,E2,E3,M1,M2,Tanwi,BCP1,BCP2}, (ii) the bouncing universe\cite{Sahni,Wands,Finelli,Martin,Khoury,Brandenberger,Tolley,Tripathy1,Tripathy2} and (iii) the cyclic universe scenario\cite{Sahni2,Ashtekar,Brown,Penrose,Ijjas1,Ijjas2,Bars,Steinhardt}. All the three scenarios have been studied extensively in the framework of different gravitational models in the past few years. Also, the dark energy and dark matter components do not fit naturally into the scheme of things in the standard big bang cosmology and have to be considered as additional contributions to the energy density of the universe. This problem can also be overcome by modifying GR\cite{DDG,fR,fRT,SPAL} or modifiying the matter source\cite{CG,MCG,BCPHDE,Zlatev,Steinhardt2,Picon,Chiba,Hu,Sahni3} in the Einstein Field Equation (EFE).   

The EU scenario was first proposed by Ellis and Maartens as an attempt to investigate whether the inflationary mechanism is past eternal\cite{EM}. As suggested by observations, the universe is almost flat at recent times\cite{Efstathiou}. However, they argued that there is always a possibility that the curvature term in the Friedmann equation had a significant contribution in the very early epochs of evolution of the universe. Considering the possibility of a closed universe in the standard relativistic context that is dominated by a minimally coupled scalar field having a physically justified potential, they found that the inflationary epoch emerges out of an Einstein static universe (ESU). The  inflationary epoch is followed by a phase of reheating that gives the standard cosmological evolution. Moreover, if the radius of the ESU does not fall below the Planck scale, then the necessity of a Quantum gravity (QG) era can be bypassed.  An EU can also be realized by modifying the Lagrangian of the Einstein-Hilbert action through addition of a term quadratic in scalar curvature with a negative coupling parameter for a particular form of the scalar field\cite{M1}. The EU scenario was first realized for a flat universe in the context of the semi-classical Starobinsky model\cite{Starobinsky}. Later on, it was also found to be viable for a standard relatistvic flat universe with the matter source described by a particular non-linear Equation of state (EoS)\cite{M2}.  

Recently, bounce models have also attracted a lot of attention. In the bounce models, the contracting universe bounces back to a re-expanding phase without reaching a singularity. There are two ways to achieve a cosmological bounce. Within the domain of standard GR, if we consider vanishing spatial curvature, the bounce can be achieved only if the null energy condition (NEC) can be violated by the matter source. Such bounce models provide an alternative to the inflationary mechanism. Alternatively, if the model assumes a positive spatial curvature, then the violation of NEC is not essential but the violation of the strong energy condition is sufficient to obatin a bounce. However, after such a bounce, there might exist a remnant spatial curvature in the expanding phase, which will again require an inflationary mechanism to dilute it, in order to make it compatible with the observations indicating towards a flat universe at present time. The second way to achieve a bounce revolves around the idea that as the contracting phase tends towards reaching a singular point, the spatial curvature and energy densities are high enough to invoke QG effects. As soon as QG effects become significant, GR being a complete classical theory of gravitation, breaks down. So high energy modifications to GR like extra dimensions and branes\cite{RS1,RS2} or loop quantum gravity (LQG)\cite{Rovelli,Bojowald} is required to figure out the transition through the bounce. There is msotly a mechanism involving a massive scalar field that causes the amplitude of each consecutive cycle of non-singular bounce to increase, resulting in an automatic resolution of the flatness problem without the inaflationary mechanism. In the cyclic scenario, besides having a bounce mechanism for averting the singularity at high matter densities, an additional cosmological turnaround mechanism is required in order to induce the contracting phase at low matter densities. As already discussed, the former mechanism is achieved either by violating the energy conditions in a standard relativistic context, or by introducing the high curvature corrections to the standard Einstein-Hilbert (EH) action. For inducing the turnaround mechanism in a relativistic context, the universe must contain some matter source which evolves with negative energy density at late times, besides ordinary matter. Such a possible matter source may be charecterized by the relation between energy density and scale factor of the form $\rho=-\frac{A}{a^n}$, where $A>0$ and $n\leq2$\cite{Sahni2}. A special case having $n=0$ corresponds to a negative cosmological constant. Also, scalar fields with certain particular form of potentials like that of a cosine one admit negative potential values in an expanding universe and can induce a turnaround. It is to be specially noted that in the context of certain models involving high curvature corrections to the EH action (both with and without extra dimensions), a quadratic correction term in the energy density of the modified effective stress energy may be significant, provided the density of a phantom dark energy is considered, which increases with the expansions of the universe. Thus the bounce results from normal matter due to high-curvature corrections and the turnaround results from phantom nature of dark energy.    

The idea of extra dimensions has always attracted attention in physics. An additional fifth dimension was first used independently by Kaluza\cite{Kaluza} and Klein\cite{Klein} to unify gravity and electomagnetic forces. The idea of extra dimensions have gained huge interest after the formulation of the Superstring/M-theories\cite{Polchinski1,Polchinski2}. The braneworld gravity\cite{RS1,RS2} is an effective theory of gravitation inspired from Supersting/M theories which reduce to standard GR in the low-energy limit. The idea is that our universe is a (3+1)-dimensional hypersurface called the 'brane' embedded in a higher dimensional spacetime known as the 'bulk'. The standard model forces are confined to the brane, while gravity is free to propagate in the bulk. In the Randall-Sundrum (RS) single brane model\cite{RS2}, which is of particular interest in the cosmological context\cite{Binetruy,Maeda,Langlois,Chen,Kiritsis,Campos,Sengupta1,Maartens,Paul,Sengupta}, the extra dimension is spacelike, characterized by a bulk space of Lorentzian signature. The brane tension is positive and the bulk space is Anti-deSitter ($AdS_5$), sourced by a negative cosmological constant. There is a fine-tuning of the brane tension and the bulk cosmological constant, leading to vanishing of the effective cosmological constant on the brane. This model is well known to modify the cosmological behaviour at early times.     

As already mentioned, LQG is a theory of quantum gravity that provides an alternative description of spacetime when the energy densities and curvature are extremely high. Such a situation arises in the cosmological context out of geodesic incompleteness as we extrapolate the evolution of the universe backward in time. Loop Quantum Cosmology (LQC) is a cosmological framework obtained by applying LQG to homogeneous systems\cite{Ashtekar2}. It is found to be significant in modifying the initial singularity and early universe scenario by introducing the central effects of LQG into effective classical equations, avoiding problems arising from quantum mechanical interpretations. It also attempts to understand more fundamental issues like understanding the geometrical structure of the spacetime surrounding the classical singularity and also, the nature of time itself. Interestingly, an extra dimensional braneworld scenario dual to the RS braneworld\cite{RS2} was proposed by Shtanov and Sahni\cite{Sahni}, which basically has a timelike extra dimension, modifying the bulk signature. It is even more interesting to note that, the modification to the standard EFE arising in this framework due to the modified action is exactly identical to modification present in LQC for a spatially flat universe.

An explicit potential for a scalar field describing a consistent EU scenario was obtained by Ellis, Murugan and Tsagas\cite{E2}. An exponential potential of the form $V(\phi)=A\big(e^{B\phi}-1\big)^2$ is found to be consistent in describing an inflationary emergent cosmology for a spatially closed universe in the standard relativistic context. However, the matter sector sourced by such a potential in standard GR can equivalently be described by modifying the EH action with an additional term quadratic in scalar curvature, identical to Starobinsky's inflationary model. Scalar fields with hyperbolic potentials finds a wide range of application in cosmology. As for example, a scalar field with a cosh potential\cite{Sahni3} which behaves like a negative exponential potential early and a power law potential at late times, can give rise to pressureless dark matter or quintessential dark energy at once the field rolls down the potential and starts oscillating rapidly about a zero field value. A potential of the form $V(\phi)=V_0 tanh^2 (\lambda_0 \phi)$ can also successfully describe an emergent cosmology in a similar spatially closed relativistic context. However, as already discussed a generalized non-linear equation of state (EoS) of the form $p=A\rho-B\rho^\frac{1}{2}$ can successfully describe an EU scenario in a relativistic context even for a spatially flat universe. Since observations favour a spatially flat universe, we would like to check whether the formation of an EU is consistent for vanishing satial curvature in a modified gravity context with and without extra dimensions. For this purpose we choose the RS-2 braneworld model and the LQC, which however may be realized as a braneworld model with a timelike extra dimension, when the spatial curvature $k=0$. However, we would like to consider it as the LQC model here to extend our analysis in the usual (3+1)-dimensions as well. The motivation behind choosing these frameworks is that they contain high energy correction terms in the field equations arising due to modification of the action from extra dimension or due to quantum effects. We shall check the evolution of the scale factor for each of the three cases in both the frameworks and if an inflationary emergent scenario arises, we shall evaluate the slow roll parameters and the mass limit on the scalar field required to begin the inflationary phase from the number of e-foldings to ensure the viability of the inflationary scenario. In the concluding section, we discuuss the physical aspects of our obtained results.

\section{Mathematical formulation}

In this section, we shall consider both the higher dimensional RS II braneworld model and the effective LQC model to study the evolution of the scale factor for the three different cases: two involving specific choice of potentials for minimally coupled scalar fields capable of reproducing emergent cosmology for spatially closed universes in the standard relativistic context and one involving matter source described by an exotic EoS that gives an emergent cosmology without involving scalar field inflation for spatially flat universes in a relativistic context. We first begin by studying the three different cases for the extra dimensional RS-II braneworld model followed by the effective LQC model.

\subsection{RS-II model}

The modified Friedmann equation on the RS II brane has the form\cite{Shiromizu}  

\begin{equation} \label{eq1}
	G_{\mu\nu}= -\Lambda g_{\mu \nu}+\kappa^2 T_{\mu\nu}+\frac{6\kappa^2}{\lambda} S_{\mu\nu}-E_{\mu\nu},
\end{equation}
where $\lambda$ is the brane tension, which is positive for our concerned braneworld model and $\kappa^2=8\pi G$. $\Lambda$ denotes the effective cosmological constant on the brane which is defined as

\begin{equation}
	\Lambda= \frac{|\Lambda_{5}|}{2}\bigg[{\bigg(\frac{\lambda}{\lambda_c}\bigg)}^2-1\bigg].
\end{equation} 

Here, $\Lambda_5$ denotes the bulk cosmological constant. As the RS II bulk spacetime ($AdS_{5}$ in nature) is characterized by a negative cosmolgical constant, so we take the modulus to take only the magnitude of it. $\lambda_c$ denotes the critical brane tension, the value of the brane tension for which there is a fine-tuning between the negative bulk cosmological constant and the positive brane tension, resulting in vanishing of the effective cosmological constant on the brane. We shall assume $\lambda=\lambda_c$ from here on to simplify the obtained solutions as the effective cosmological constant on the brane plays no significant role in our analysis and this assumption was considered by RS in their model too.

The term $S_{\mu \nu}$ is a correction term in the modified field equation that represents the local correction on the 3-brane consisting of terms quadratic in stress-energy. It has the form
\begin{equation} \label{eq2}
	S_{\mu\nu}= \frac{TT_{\mu\nu}}{12}-\frac{T_{\mu\alpha}T_{\nu}^{\alpha}}{4}+ \frac{g_{\mu\nu}}{24}(3T_{\alpha\beta}T^{\alpha\beta}-T^2),
\end{equation}
where $T$ denotes trace of the stress-energy tensor.

The term $E_{\mu\nu}$ is another correction term in the modified Einstein Field Equation (EFE) that represents the projected Weyl tensor on the brane computed at the brane and encodes the bulk gravitational effects on the brane. So, it is the non-local correction to the EFE. A unit vector $n^A$ is defined normal to the brane, such that $g^{AB} = g^{AB}_{(5)}- n^An^B$. We have $E_{\mu\nu} = C^{(5)}_{ACBD}n^C n^D g^A_{\mu} g^B_{\nu}$.

The modified Friedmann equation on the brane can now be written as 
\begin{equation}
	H^2 = \frac{\Lambda}{3}+\frac{\kappa^2}{3}\rho \left(1 + \frac{\rho}{2\lambda}\right) +\frac{C}{a^4},
\end{equation}
where we have already set the effective cosmological constant $\Lambda=0$ as $\lambda=\lambda_c$ following RS. The parameter $C$ is a constant of integration and the $\frac{C}{a^4}$ term arises from the contribution of the projected Weyl tensor on the brane. Physically, it represents a form of radiation as the projected Weyl tensor on the brane is traceless and is often called the dark radiation. However, this term can be ignored in the cosmological evolution as the constant of integration $C$ can be set to zero\cite{Binetruy}.  

The conservation equation holds true separately on the 3-brane and the higher dimensional bulk spacetime. So, on the brane, the conservation of stress-energy yields the same equation as in the case of standard relativistic cosmology

\begin{equation}
	\dot{\rho}+3H(p+\rho)=0.
\end{equation} 

\underline{CASE 1 (Non-linear Equation of state)}

We consider an EoS for a perfect fluid that is capable of describing presureless non-baryonic cold dark matter, exotic matter as well as dark energy. Such an EoS was used by Mukherjee et al.\cite{M2} to realize an EU scenario without any inflationary mechanism involving a minimally coupled scalar field. The  spatially flat universe was found to emergy to its present state from an Einstein static universe (ESU). The EoS is given as 
\begin{equation}
	P = A\rho - B\rho^{1/2}
	\label{state}
\end{equation}

Using this EoS in the conservation Eq. (5), we immediately have
\begin{equation}
	\dot{\rho} + 3H\left[\left(A+1\right)\rho - B\rho^{(1/2)}\right] = 0 
	\label{eq.state}
\end{equation}

On solving the above ordinary differential equation, one obtains a solution for the energy density $\rho$ in terms of the scale factor $a(t)$ and the constant EoS parameters $A$ and $B$ as
\begin{equation}
	\rho = \left[\frac{\left(\frac{C}{a^3}\right)^\frac{A+1}{2} + B}{A+1}\right]^2,
	\label{sol_rho}
\end{equation}
where $C$ is a constant of integration.

Using the modified Friedmann equation in the early universe, we obtain a differential equation, integrating both sides of which we obtain 

\begin{equation}
	\int \frac{da}{a\left[\frac{\alpha}{a^{3(A+1)/2}} + \beta\right]^2} = \int\frac{\kappa}{\sqrt{6\lambda}}(A+1)^{-2}dt,
	\label{integral_RS}
\end{equation}

where we introduce the constant parameters $\alpha = \frac{C^{(A+1)/2}}{A+1}$ and $\beta = \frac{B}{A+1}$.

On solving the above equation and assuming that we are concerned with the early universe scenario only, we get

\begin{equation}
	\left[1 + \frac{\beta}{\alpha}a^{3(A+1)/2}\right]^{\frac{1}{\beta^2}} = e^{\left[\left(3/2(A+1) + 1\right)\left(\frac{\kappa}{\sqrt{6\lambda}}t + C^{\prime \prime}\right) + a_0\right]},
	\label{sol_scale2}
\end{equation}
where $a_0$ represents a small non-zero constant and $C^{\prime \prime}$ is an integration constant.

From this equation we can obtain an expression for the scale factor $a(t)$, given by 

\begin{equation}
	a = \left[\frac{\alpha}{\beta}\left[e^{\left[\left(3/2(A+1) + 1\right)\left(\frac{\kappa}{\sqrt{6\lambda}}t + C^{\prime \prime}\right) + a_0\right]^{\beta^2}} - 1\right]\right]^{\frac{2}{3(A+1)}}
	\label{scale_eq._state_RS}
\end{equation}

We plot the variation of the scale factor with time in Figure (1). As evident from the plot, the scale factor reaches a zero value in the finite past and we obtain an universe which is geodesically incomplete in the past. As a result, the EU scenario can not be realized for the exotic EoS in the braneworld scenario for a spatially flat universe although it works well in a standard relativistic context. The probable reason for this is discussed in the concluding section. 
\begin{figure}
	\centering
	\includegraphics[width=0.5\textwidth]{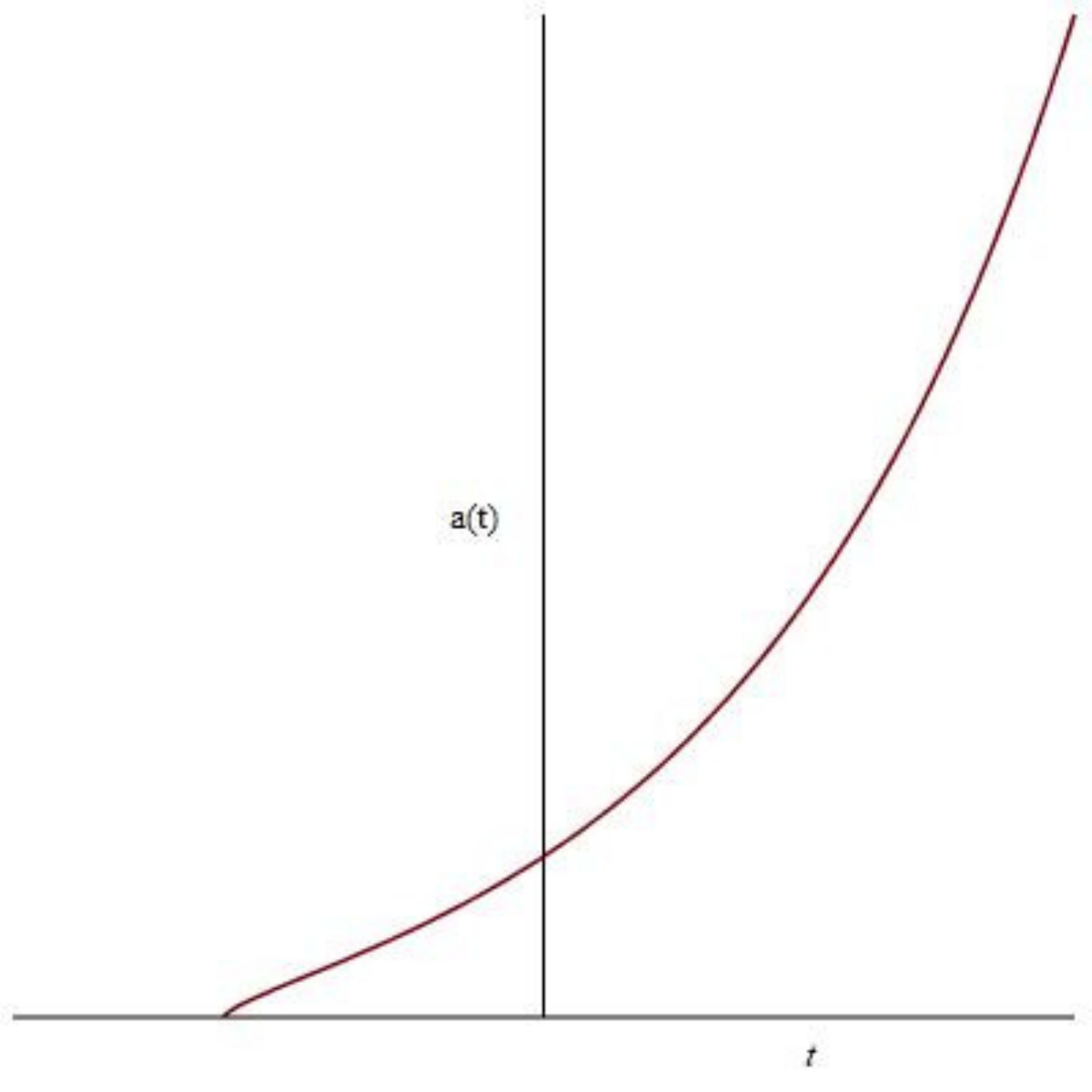}
	\caption{Variation of scale factor a(t) with time t for Case-1 of RS-II braneworld.}\label{RS3}
\end{figure}

\underline{CASE 2 (Scalar field with exponential potential)}

The action for a minimally coupled scalar field may be written as
\begin{equation}
	S=-\int{\bigg(\frac{1}{2}g^{\mu \nu}\partial_\mu \phi \partial_\nu \phi + V(\phi)\bigg)\sqrt{-g}d^4x}
\end{equation}

The Euler-Lagrange equation of motion for the scalar field can be obtained from the above Lagrangian density as 
\begin{equation}
	\ddot \phi + 3H\dot\phi + \frac{dV}{d\phi} = 0
\end{equation}

This equation can also be arrived at from the conservation equation (5) once we plug in the components of the stress-energy tensor for the scalar field. The energy momentum tensor can be obtained from the action using the definition $T_{\mu \nu}=-\frac{2}{\sqrt{-g}}\frac{\delta S}{\delta g^{\mu \nu}}$, which yields

\begin{equation}
	T_{\mu \nu}=\partial_{\mu} \phi \partial_{\nu} \phi-g_{\mu \nu} \bigg[\frac{1}{2}g^{\mu \nu}\partial_\mu \phi \partial_\nu \phi + V(\phi)\bigg]
\end{equation}

\begin{figure}
	\centering
	\includegraphics[width=0.5\textwidth]{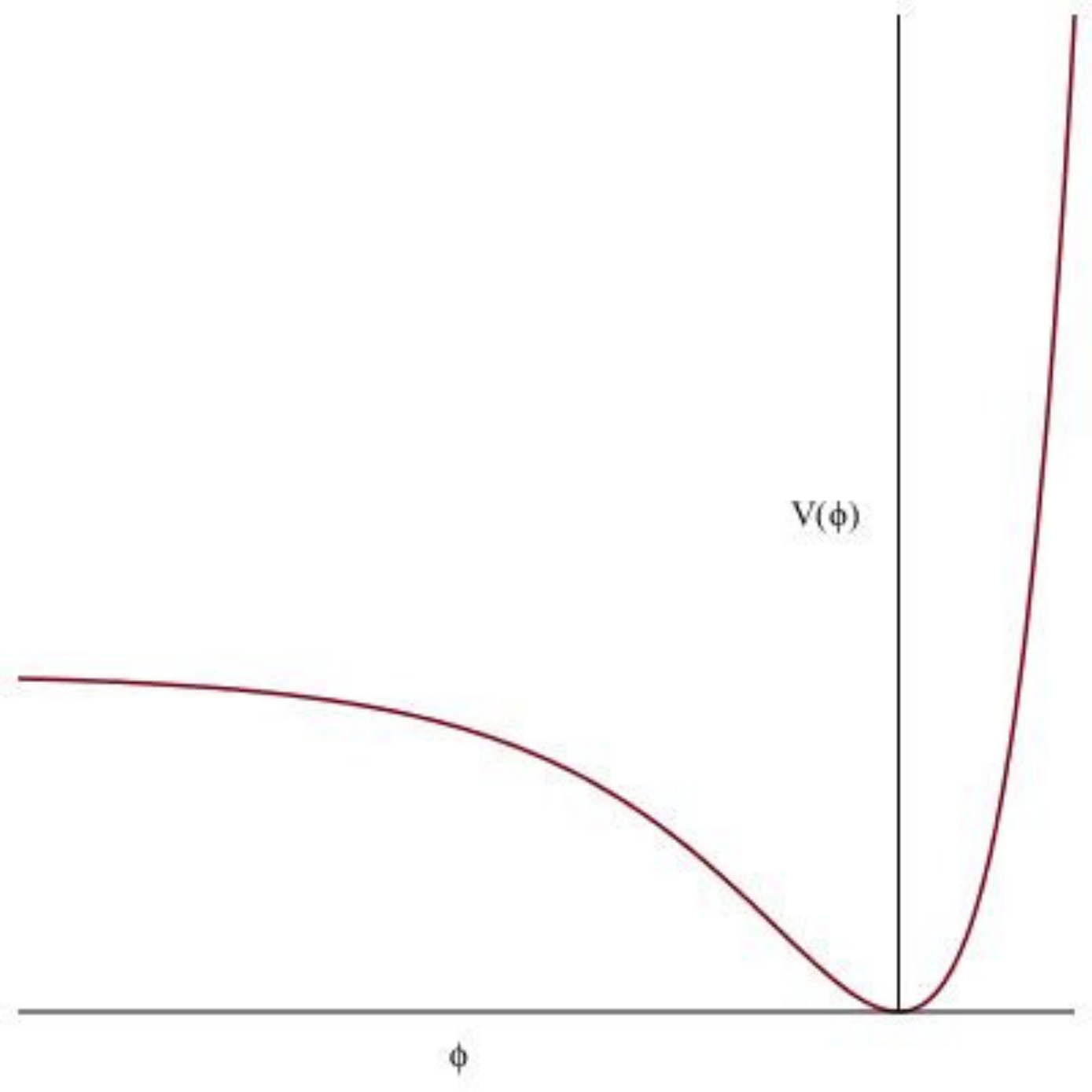}
	\caption{Variation of scalar field potential $V(\phi)$ with field $\phi$ for first potential.}\label{POTENTIAL1}
\end{figure}

Assuming isotropy and homogeneity, the energy density and pressure of the scalar field can be obtained as
\begin{equation}
	\rho=\frac{\dot{\phi}^2}{2}+V(\phi),
\end{equation}
and \begin{equation}
	p=\frac{\dot{\phi}^2}{2}-V(\phi)
\end{equation}

We shall consider a weakly decelerated scalar field implying that $|\ddot{\phi}|<|\frac{dV}{d\phi}|$, which simplifies the equation of motion or the conservation equation for the scalar field to

\begin{equation}
	3H\dot\phi + \frac{dV}{d\phi} = 0
	\label{conservation}
\end{equation}

Now we consider the first potential for the scalar field, for which we want to obtain the scale factor in a spatially flat universe on the brane $V(\phi)=A{(e^{B \phi}-1)}^2$, where $A$ and $B$ are constant parameters. The variation of the potential with the scalar field has been plotted in Figure 2.

Using the modified Friedmann equation on the brane and Eq. (18) with the above mentioned form of the potential, we obtain $\dot{\phi}$ as

\begin{equation}
\dot{\phi} = \frac{2}{3}\frac{B \sqrt{6\lambda}}{\kappa}\left(1 - e^{-B\phi(t)}\right)^{-1}
\label{phi_derivative_first_pot_RS}
\end{equation}
We are interested in checking whether an EU scenario can be realized and whether there is any geodesic incompleteness in the past. So, the quadratic correction term in the modified Friedmann equation dominates as we deal with high energy scales. For lower energy scales the brane tension is large resulting in redundant significance of the correction term and standard GR is recovered. 

On integrating we get
\begin{equation}
\phi + \frac{e^{-B\phi}}{B} = \frac{2}{3}\frac{B\sqrt{6\lambda}}{\kappa}t + C_1,
\label{phi_first_pot_RS}
\end{equation}
where $C_1$ is the constant of integration.

The field $\phi$ turns out to have the form

\begin{equation}
\phi(t) = \frac{1}{3}\frac{2B^2\sqrt{6\lambda}t + 3BC_1\kappa + 3LambertW\left(-e^{-\frac{2B^2\sqrt{6\lambda}t - 3BC_1\kappa}{3\kappa}}\right)\kappa}{B\kappa}
\label{phi_final_first_pot_RS}
\end{equation}

As we are concerned with the early universe, so the above expression can be reasonably approximated to have the following form

\begin{equation}
\phi(t) = \frac{1}{3}\frac{2B^2\sqrt{6\lambda}t + 3BC_1\kappa + 3LambertW\left(\frac{2B^2\sqrt{6\lambda}t - 3BC_1\kappa}{3\kappa}-1\right)\kappa}{B\kappa}
\label{phi_final_first_pot_RS_simplified}
\end{equation}

The variation of the obtained field with time has been plotted in Figure 3.
\begin{figure}
\centering
\includegraphics[width=0.5\textwidth]{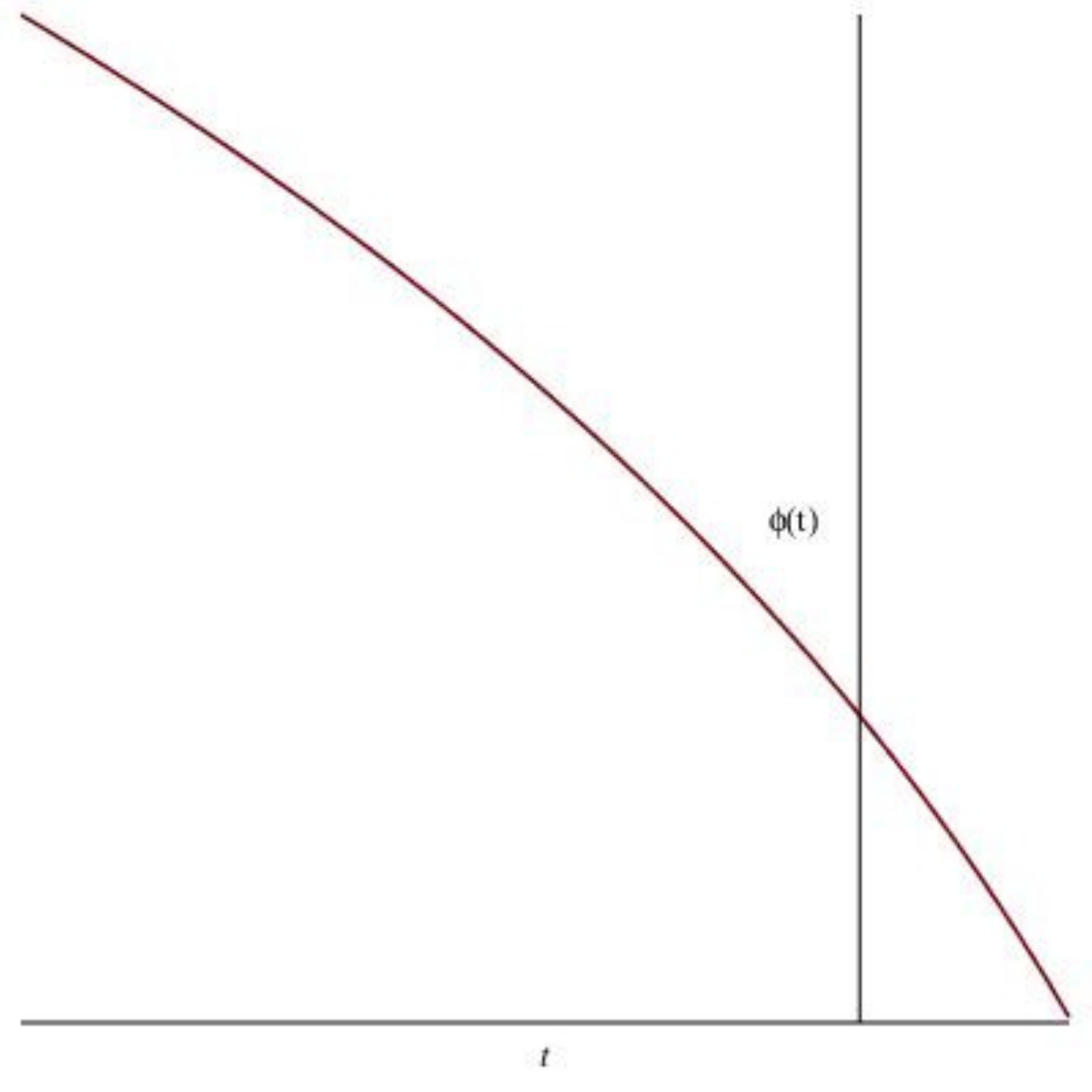}
\caption{Evolution of scalar field $\phi(t)$ with time($t$) for case-2 of RS-II braneworld.}\label{RS-FIELD1}
\end{figure}

Plugging in $\dot{\phi}$ from Eq.(19) into the simplified equation of motion for the scalar field and using the expression of the Hubble parameter as $H=\frac{\dot{a}}{a}$, the scale factor in terms of the field for our particular choice of the scalar field potential turns out to be

\begin{equation}
a = C_2e^{-\frac{1}{4}\frac{A \kappa^2}{B\lambda}\left[\frac{e^{2B \phi}}{2B} - \phi - 3\frac{e^{B \phi}}{B} + \frac{e^{-B \phi}}{B}\right]},
\label{scale_first_pot_RS}
\end{equation}
where $C_2$ denotes an integration constant.

Finally, plugging in the obtained field in this equation, we get the scale factor as

\begin{multline}
	a = C_2e^{-\frac{1}{4}\frac{A\kappa^2}{B\lambda}}\left[\frac{e^{\frac{4B^2\sqrt{6\lambda}t + 3BC_1\kappa + 3LambertW\left(\frac{2B^2\sqrt{6\lambda}t - 3BC_1\kappa}{3\kappa} -1\right)\kappa}{3\kappa}}}{2B}\right.
	\\ \left.-\frac{1}{3}\frac{2B^2\sqrt{6\lambda}t + 3BC_1\kappa + 3LambertW\left(\frac{2B^2\sqrt{6\lambda}t - 3BC_1\kappa}{3\kappa}-1\right)\kappa}{B\kappa}\right.
	\\ \left. -3\frac{e^{\frac{2B^2\sqrt{6\lambda}t + 3BC_1\kappa + 3LambertW\left(\frac{2B^2\sqrt{6\lambda}t - 3BC_1\kappa}{3\kappa}-1\right)\kappa}{3\kappa}}}{B} + \frac{e^{-\frac{2B^2\sqrt{6\lambda}t + 3BC_1\kappa + 3LambertW\left(\frac{2B^2\sqrt{6\lambda}t - 3BC_1\kappa}{3\kappa}-1\right)\kappa}{3\kappa}}}{B} \right]
	\label{scale_first_pot_RS_time}
\end{multline}

\begin{figure}
\centering
\includegraphics[width=0.5\textwidth]{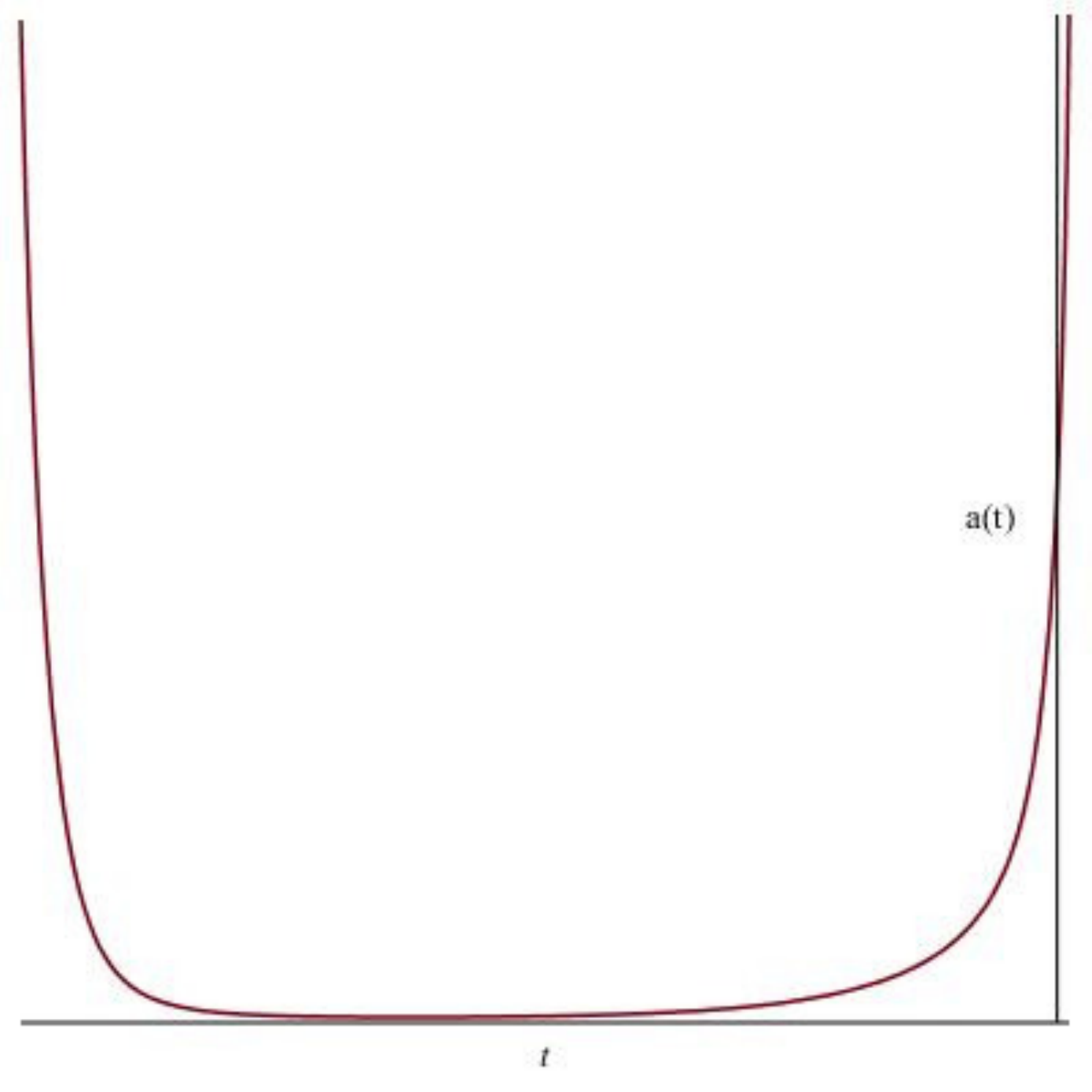}
\caption{Variation of scale factor a(t) with time t for Case-2 of RS-II braneworld.}\label{RS1}
\end{figure}

The evolution of the scale factor with time has been plotted in Figure 4. As we can see from the figure the evolution of the universe is geodesically complete in the past as no singularity is attained. There is a non-singular bounce that prevents the universe from reaching a singular state. The contraction phase is followed by a quasi-static state before the universe starts expanding again. So, the initial singularity is absent. 

\underline{CASE 3 (Scalar field with hyperbolic potential)}

Next we consider a minimally coupled scalar field with a potential of the form $V=V_0 tanh^2(\lambda \phi)$. The evolution of the potential with the field has been plotted in Figure 5. 

\begin{figure}
\centering
\includegraphics[width=0.5\textwidth]{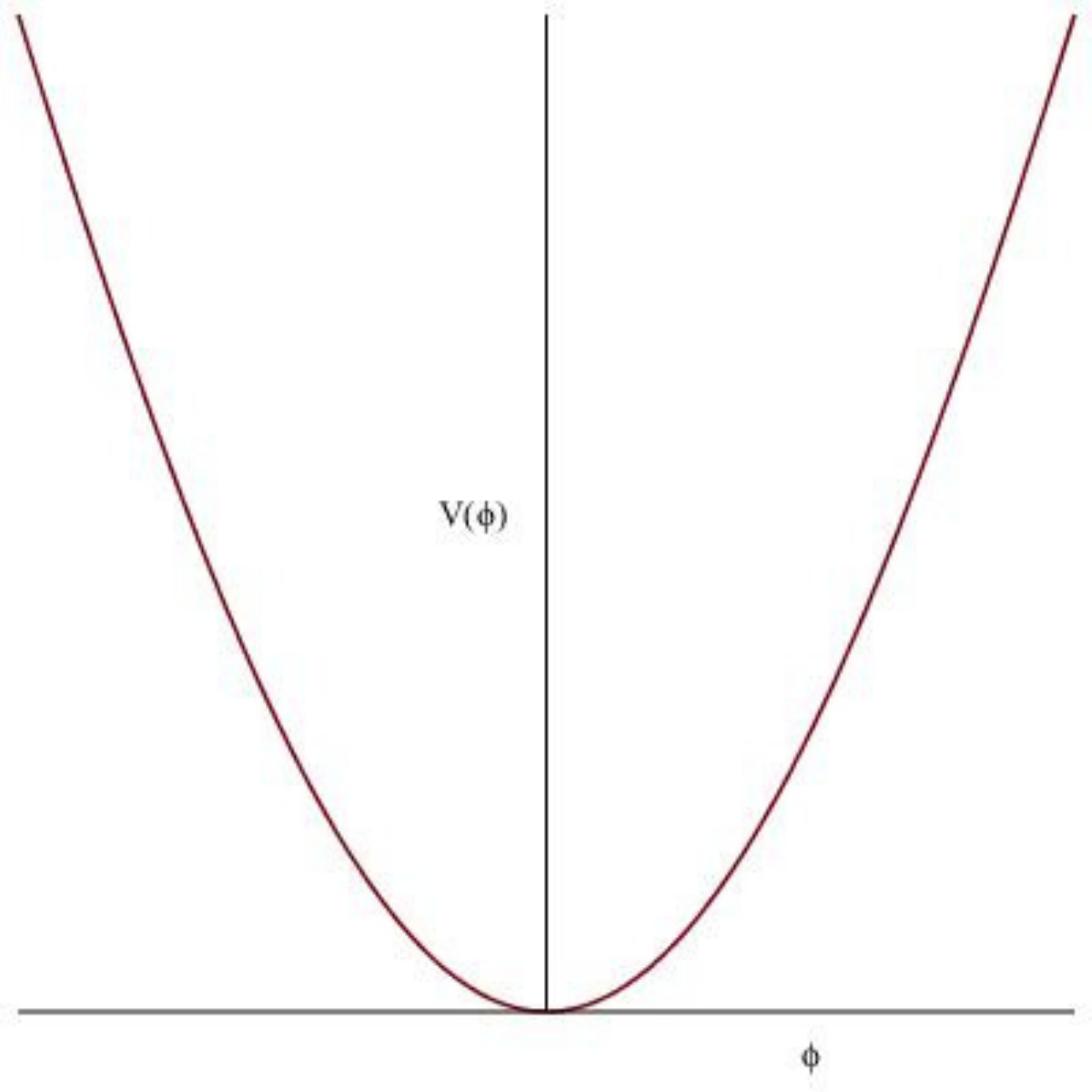}
\caption{Variation of scalar field potential $V(\phi)$ with field $\phi$ for second potential.}\label{POTENTIAL2}
\end{figure}

Just like for the previous potential, using the modified field equation on the brane, the simplified equation of motion for the scalar field and the form of the potential, we get
\begin{equation}
\dot{\phi} = \frac{4\lambda_0\sqrt{6\lambda}}{3\kappa sinh(2\lambda_0\phi)}
\label{phi_deri_second_pot_RS}
\end{equation}

Solving the above differential equation, we obtain an expression for the scalar field $\phi(t)$ as
\begin{equation}
\phi(t) = \frac{1}{2\lambda_0}arccosh\left(\frac{8\sqrt{6\lambda}\lambda_0^2\left(t-C_3)\right)}{3\kappa}\right),
\label{phi_final_second_pot_RS}
\end{equation}
$C_3$ denoting a constant of integration.

The time evolution of the obtained scalar field has been plotted in Figure 6.
\begin{figure}
\centering
\includegraphics[width=0.5\textwidth]{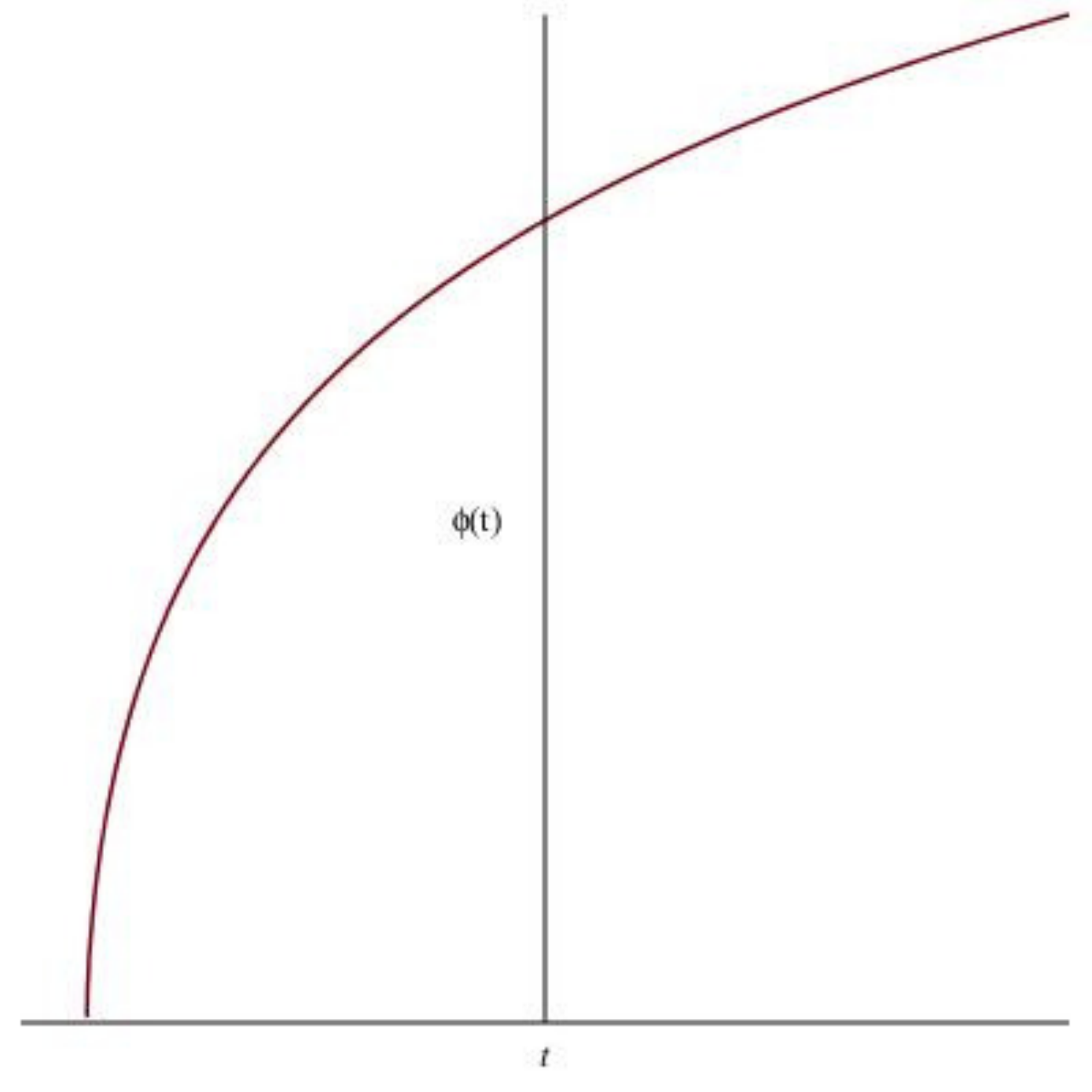}
\caption{Evolution of scalar field $\phi(t)$ with time($t$) for case-3 of RS-II braneworld.}\label{RS-FIELD2}
\end{figure}

Plugging in the expression for $\dot{\phi}$ and replacing the Hubble parameter in terms of the scale factor in the simplified conservation equation for the scalar field, we obtain the scale factor to have the form 

\begin{equation}
a = C_4e^{-\frac{9}{144}\frac{V_0 \kappa^2}{\lambda_0\lambda}\left[\frac{e^{2\lambda_0 \phi}}{2\lambda_0} + 4\phi + \frac{e^{-2\lambda_0 \phi}}{2\lambda_0} - \frac{4 ln(e^{2\lambda_0 \phi}+1)}{\lambda_0}\right]}
\label{scale_second_pot_RS}
\end{equation}
Here, $C_4$ denotes the integration constant.

Using the obtained value of the field in the above obtained expression, the scale factor as a function of time can be realized to have the form 

\begin{multline}
	a = C_4e^{\frac{-9}{144}\frac{V_0\kappa^2}{\lambda_0^2\lambda}\left[\frac{e^{arccosh\frac{8\sqrt{6\lambda}t\lambda_0^2\left(t-C_3\right)}{3\kappa}}}{2} + 2arccosh\left(8\sqrt{6\lambda}t\lambda_0^2\left(t-C_3\right)\right) + \frac{e^{arccosh\frac{8\sqrt{6\lambda}t\lambda_0^2\left(t-C_3\right)}{3\kappa}}}{2}\right.}
	\\ \left.-4ln\left(e^{arccosh\frac{8\sqrt{6\lambda}t\lambda_0^2\left(t-C_3\right)}{3\kappa}}+1\right)\right]
	\label{scale_second_pot_RS_time}
\end{multline}

We plot the time evolution of the scale factor in Figure 7. As evident from the plot, the scale factor evolves from a small non-zero value and the inflationary phase emerges out of a quasi-static phase. Thus an EU scenario can be successfully realized. 
\begin{figure}
\centering
\includegraphics[width=0.5\textwidth]{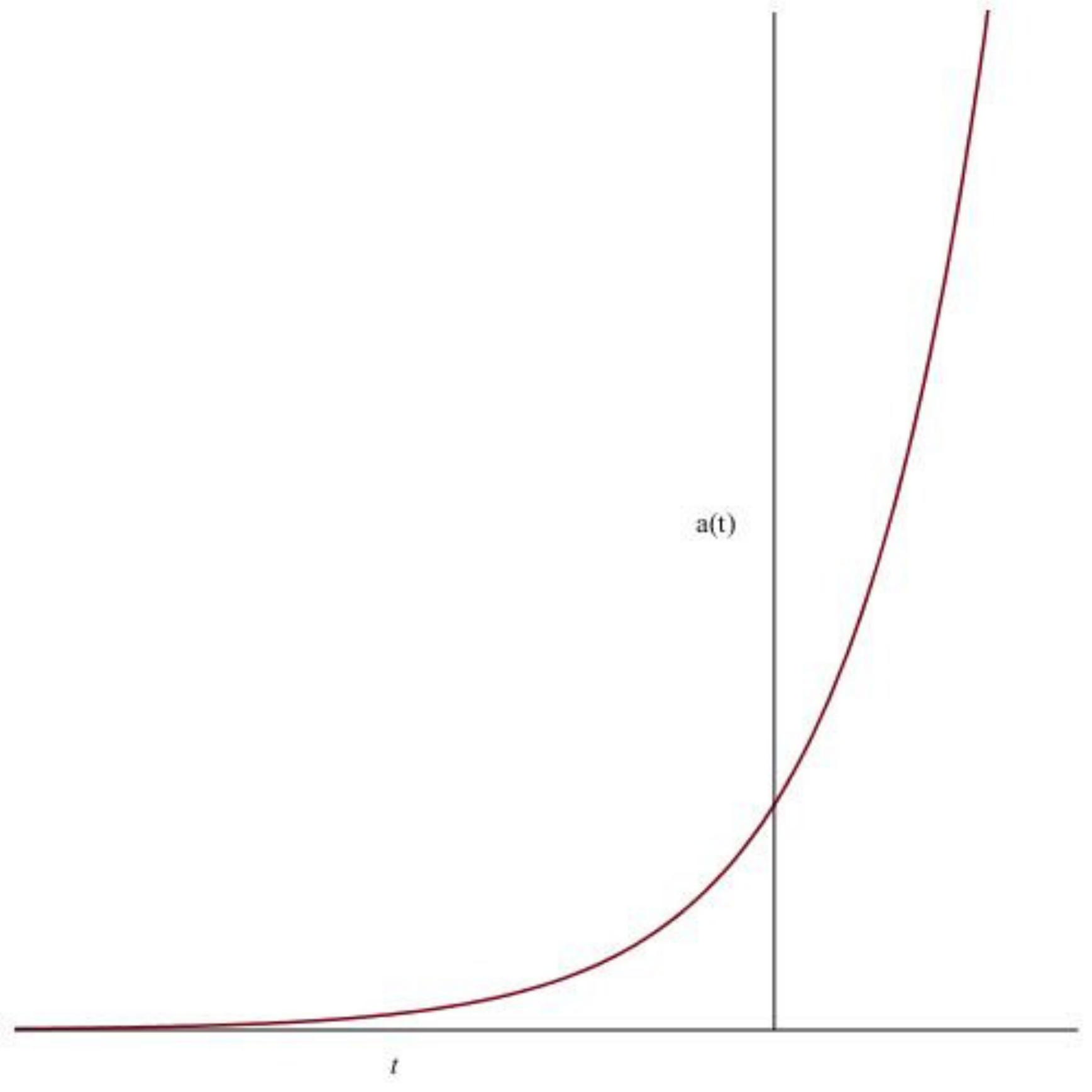}
\caption{Variation of scale factor a(t) with time t for Case-3 of RS-II braneworld}\label{RS2}
\end{figure}

We shall now evaluate the slow-roll parameters on the brane for this potential and also find out the minimum value of the scalar fiel mass that is required to begin the inflationary phase by considering more than 60 e-folds of inflation to be sufficient.

The slow-roll parameters for RS braneworld are given by,

\begin{equation}
\epsilon= \frac{V^{\prime2}}{2V^2} \frac{1+\frac{V}{\mu^4}}{\left(1+\frac{V}{2\mu^4}\right)^2} \frac{1}{\lambda_p^2}
\end{equation}

\begin{equation}
\eta=\frac{1}{\lambda_p^2}\frac{V^{\prime\prime}}{V}\left[\frac{1}{1-\frac{V}{2\mu^4}}\right]
\end{equation}

(where $\mu^4=\lambda$, $\lambda_p^2=\kappa^2$).

For our chosen form of the hyperbolic potential, the slow roll parameters on the brane have the form,

\begin{equation}
\epsilon=\frac{16\lambda_0^2\lambda}{\lambda_p^2} \frac{sech^2(\lambda_0\phi)}{tanh^2(\lambda_0\phi)}\left[\frac{\lambda + V_0tanh^2(\lambda_0\phi)}{\left(2\lambda +V_0tanh^2(\lambda_0\phi)\right)^2}\right]
\label{epsilon_RS}
\end{equation}

\begin{equation} 
\eta=\frac{1}{\lambda_p^2}\left[\frac{2\lambda_0^2V_0sech^2(\lambda_0\phi)\left(sech^2(\lambda_0\phi)-2tanh^2(\lambda_0\phi)\right)}{V_0tanh^2(\lambda_0\phi)}\right]\left[\frac{2\lambda}{2\lambda - V_0tanh^2(\lambda_0\phi)}\right]
\label{eta_RS}
\end{equation}

The number of e-folding on the RS II braneworld is given by, 

\begin{equation}
N = \lambda_p^2 \int_{\phi_f}^{\phi_i}\frac{V}{V^\prime}\left(1 + \frac{V}{2\lambda}\right) d\phi
\end{equation}

For our chosen form of the potenial, this turns out to be,

\begin{equation}
N = \frac{\lambda_p^2cosh^2(\lambda_0(\phi_i-\phi_f))}{4\lambda_0^2}\left[1 + \frac{V_0}{2\lambda}\right] - \frac{\lambda_p^2V_0}{4\lambda_0^2\lambda}ln\left(cosh(\lambda_0(\phi_i-\phi_f))\right)
\label{e-folding_RS}
\end{equation}

The lower limit on the value of $\phi_i$ can be obtained by taking the number of e-folds to be greater than 60 in order to have sufficient inflation. We find that $\phi_i$ must be greater than $34.691M_5$ to begin with. The 5D Planck mass $M_5<10^{17} GeV$ which is considerably lower than the usual Planck  mass $M_p$. So the mass value of the scalar field is lower than the Planck mass to begin with on the brane as opposed to in standard GR. Hence, we have an improved inflationary emergent scenario on the brane. 

\subsection{LQC}

In the framework of LQC, the effective quantum effects due to the quantum geometry play a significant role. The most significant effect can be said to be the introduction of a new repulsive force, the effect of which can be noted only at considerably high spacetime curvatures. In low or moderate spacetime curvatures, the effect of this force is negligeble compared to the gravitational attraction but at very high spacetime curvatures the effect of this repulsive force starts to dominate over the gravitational attraction. This results in a significant modification of the cosmological dynamics for the very early universe. In LQC any curvature invariant growing to the Planck scale can be diluted by the effective quantum geometry effects, provided the the energy conditions of standard GR are followed by the physical matter. This leads to the resolution of singularities in LQC. In LQC, the modified Friedmann equation has the form

\begin{equation}
H^2 = \frac{\kappa^2}{3}\rho \left(1 - \frac{\rho}{\rho_c}\right)
\label{friedmann1-}
\end{equation}

\underline{CASE 1 (Non-linear Equation of state)}

Like in the case of the RS-II braneworld, we first check the viability of an EU scenario with the exotic EoS given by Eq. (6). As there is no change in the conservation equation, so the nergy density can be obtained identical to Eq. (8) bu plugiing in the EoS in the conservation equation and then solving the differential equation for $\rho$. 

Using the expression for $\rho$ in the modified Friedmann equation, we arrive at the differential equation
\begin{equation}
\frac{\dot{a}}{a}=\sqrt{\frac{1}{3(A+1)^2}}\frac{1}{(aC)^{\frac{3(A+1)}{2}}}\bigg[1-\frac{\rho_c(aC)^{3(A+1)}}{2}-2B\rho_c(aC)^{\frac{3(A+1)}{2}}\bigg],	
\end{equation}	
where $C$ is a constant of integration.

On solving the above differential equation for the scale factor $a$, we obtain
\begin{equation}
a = \left\{\frac{\sqrt{4B^2{\rho_c}^2+2\rho_c}}{\rho_c}tanh\left[\frac{\sqrt{3}(t-t_0)}{4\sqrt{4B^2{\rho_c}^2+2\rho_c}}\right]-2B\right\}^\frac{2}{3(A+1)}
\label{scale_eq._state_SS}
\end{equation}
\begin{figure}
\centering
\includegraphics[width=0.5\textwidth]{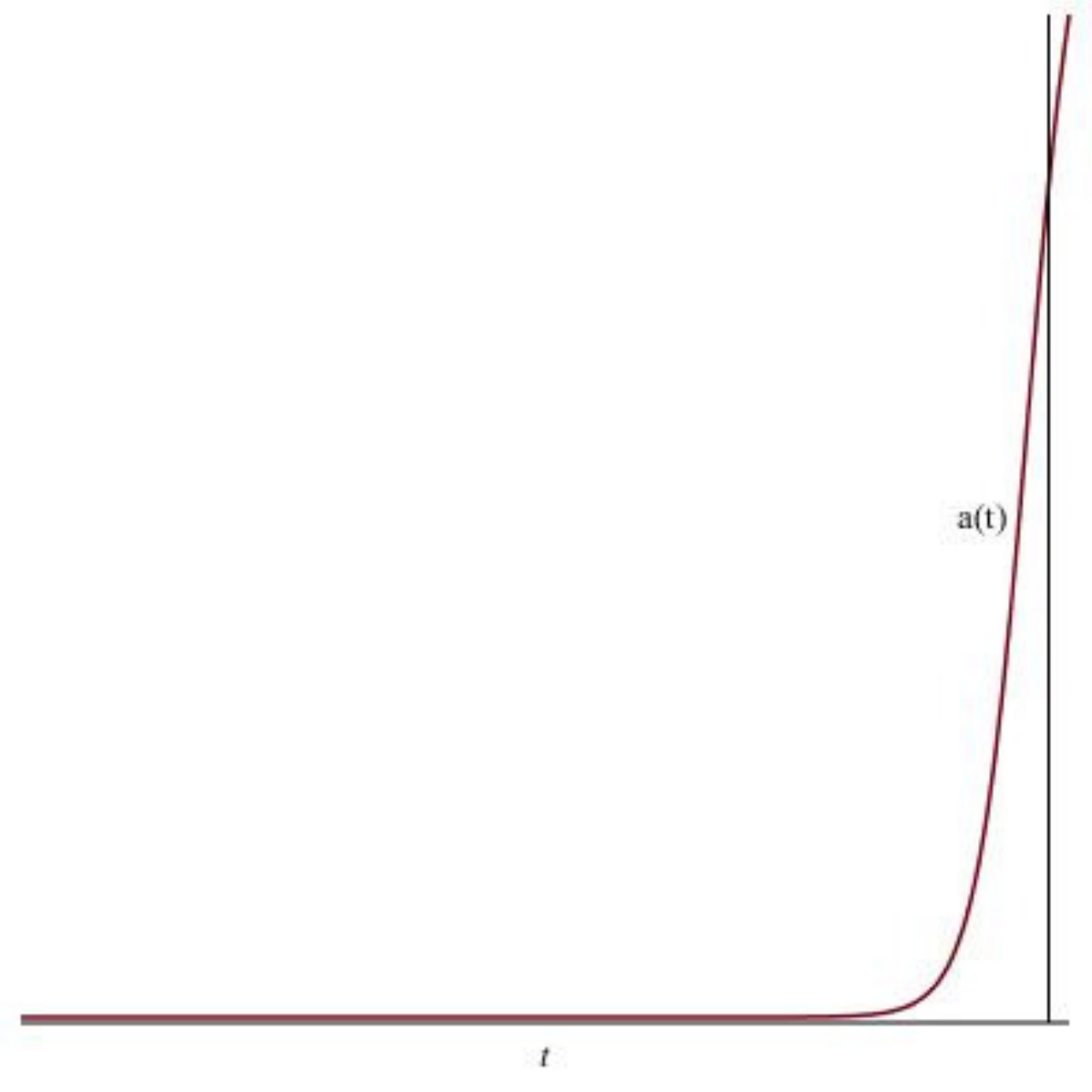}
\caption{Variation of scale factor a(t) with time t for Case-1 of LQC.}\label{SS3}
\end{figure}
The time evolution of the scale factor has been plotted in Figure 8. As evident from the plot, an EU scenario can be realized as the universe emerges to the present state from an eternal quasi-staic state followed by a phase of expansion resulting from the exotic fluid without any scalar field or cosmological constant.

\underline{CASE 2 (Scalar field with exponential potential)}

Next we move on to obtain the evolution of the scale factor for a scalar field with the potential suggested by Ellis et al. in the relativistic context for a spatially closed universe to realize an EU. Using the equation of motion for the scalar field given by Eq. (17) and the modified Friedmann equation in LQC given by Eq. (34), we obtain $\dot{\phi}$ to be described by the equation

\begin{equation}
\dot{\phi} = B\sqrt{\frac{A\rho_c}{6\pi G}}\frac{e^{B\phi}}{\sqrt{\rho_c-A(e^{B\phi-1)^2}}}
\label{phi_derivative_first_pot_SS}
\end{equation}

Solving the above differential equation, one obtains the scalar field to be given as

\begin{equation}
\phi(t)=\frac{1}{B} ln\left[\frac{Atan^2(\frac{c(t-t_0)}{\sqrt{A}})+\sqrt{\rho_c A tan^4(\frac{c(t-t_0)}{\sqrt{A}}+\rho_c A tan^2(\frac{c(t-t_0)}{\sqrt{A}}}+A}{A(tan^2(\frac{c(t-t_0)}{\sqrt{A}}+1)}\right], 
\label{phi_first_pot_SS}
\end{equation}
where $t_0$ and $c$ denote integration constants.

The time evolution of the scalar field has been plotted in Figure 9. It turns out that the field has an oscillatory nature.
\begin{figure}
\centering
\includegraphics[width=0.5\textwidth]{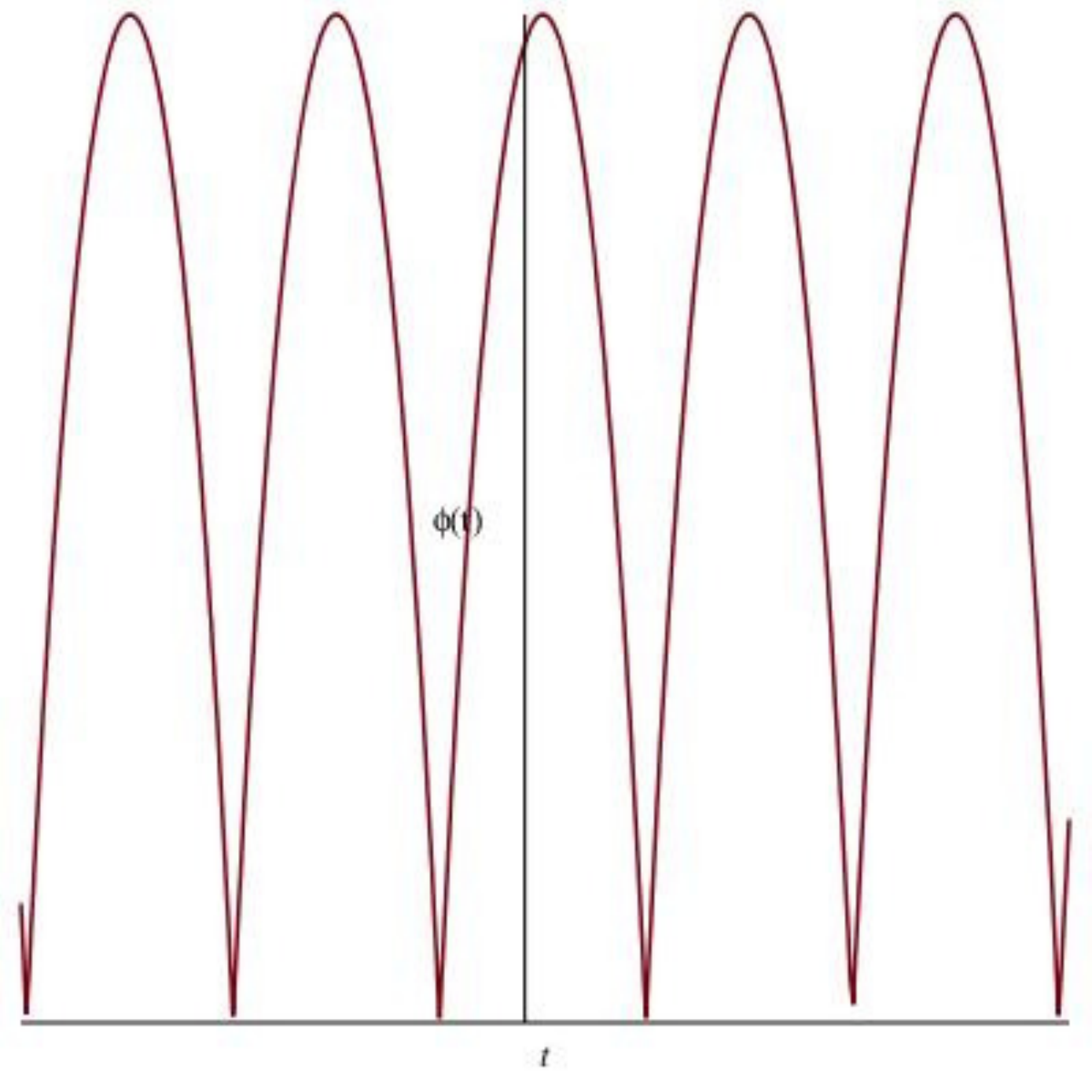}
\caption{Evolution of scalar field $\phi(t)$ with time($t$) for case-2 of LQC.}\label{SS-FIELD1}
\end{figure}

Pluging in the time derivative of the field and replacing the Hubble parameter as $H=\frac{\dot{a}}{a}$ in Eq. (17) along with the form of the potential $V(\phi)$, we obtain a differential equation for the scale factor $a$ on solving which we get

\begin{equation}
a = exp\left[-C -\frac{4\pi G}{AB}\left\{\phi+\frac{1}{Be^{B\phi}}\right\}\right],
\label{scale_first_pot_SS}
\end{equation}
where $C$ is an integration constant.

Plugging in the obtained expression for the field $\phi$, the scale factor has the form

\begin{multline}
	a = exp\left[-C -\frac{4\pi G}{AB^2}\left\{ ln\left[\frac{Atan^2(\frac{c(t-t_0)}{\sqrt{A}})+\sqrt{\rho_c A tan^4(\frac{c(t-t_0)}{\sqrt{A}}+\rho_c A tan^2(\frac{c(t-t_0)}{\sqrt{A}}}+A}{A(tan^2(\frac{c(t-t_0)}{\sqrt{A}}+1)}\right]\right.\right.
	\\ \left.\left.+\frac{1}{Be^{ln\left[\frac{Atan^2(\frac{c(t-t_0)}{\sqrt{A}})+\sqrt{\rho_c A tan^4(\frac{c(t-t_0)}{\sqrt{A}}+\rho_c A tan^2(\frac{c(t-t_0)}{\sqrt{A}}}+A}{A(tan^2(\frac{c(t-t_0)}{\sqrt{A}}+1)}\right]}}\right\}\right]
	\label{scale_first_pot_SS_time}
\end{multline}

\begin{figure}
\centering
\includegraphics[width=0.5\textwidth]{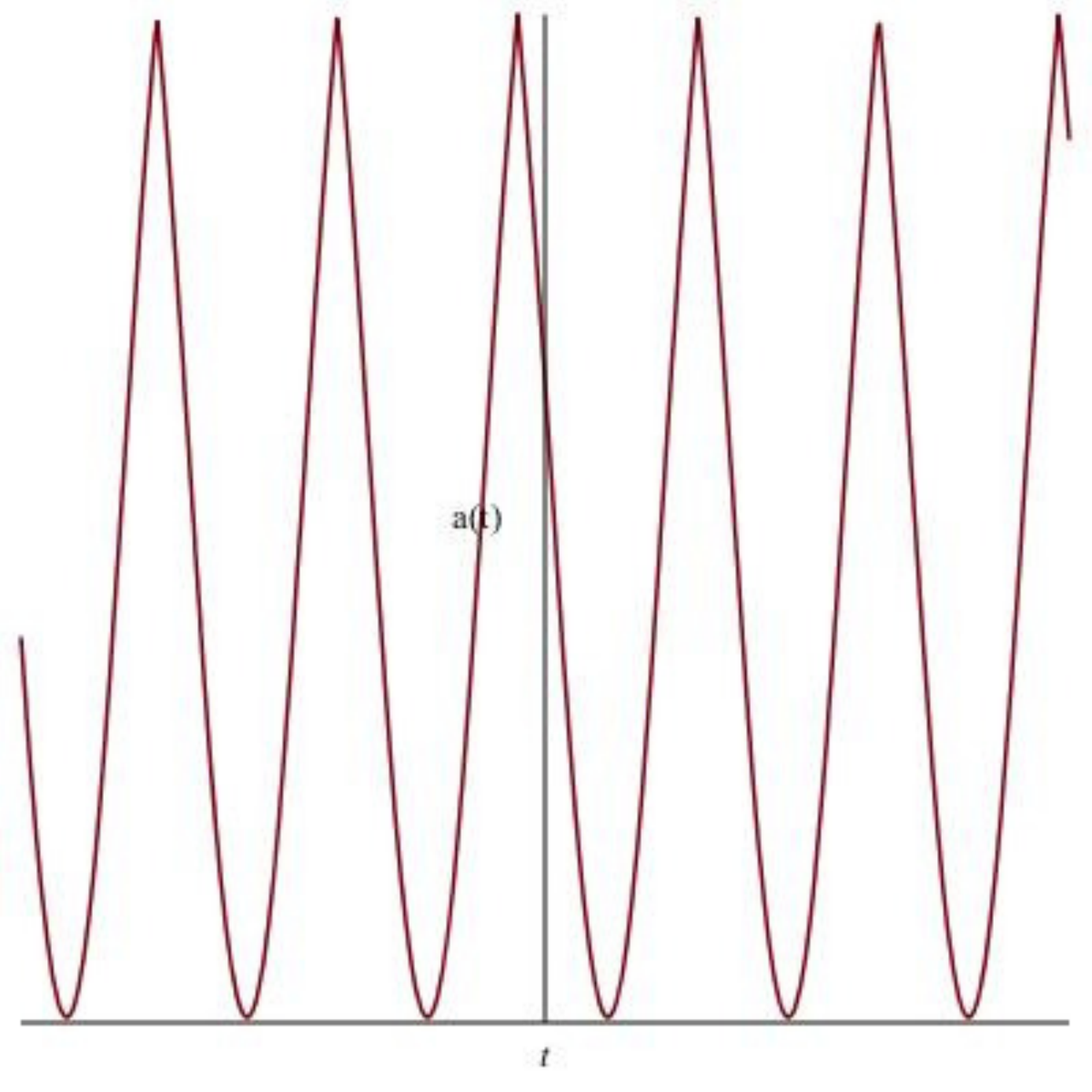}
\caption{Variation of scale factor a(t) with time t for Case-2 of LQC.}\label{SS1}
\end{figure}
The time evolution of the scale factor is plotted in Figure 10. As we can see the universe is non-singular but we obtain a cyclic universe rather than an inflationary emergent one. There is repetitive non-singular bounce following every epoch of contraction and then we have a cosmic turnaround leading the universe to an expanding phase. The cosmic turnaround is sourced by the scalar field and the singularity can be avoided due to the effect of the repulsive force that arises from the effective quantum geometry effects and takes over the gravitational attraction owing to high spacetime curvatures.

\underline{CASE 3 (Scalar field with hyperbolic potential)}

In the third and final case, we study the evolution of the scale factor with time for the potential $V(\phi)=V_0 tanh^2(\lambda_0 \phi)$. Using Eq. (17) and (34) and the mentioned form of $V(\phi)$, a differential for $\phi(t)$ acn be arrived at, having the form 

\begin{equation}
\dot{\phi} = \frac{C\lambda_0 \sqrt{\rho_c}}{\sqrt{6\pi G}}\frac{ sech^4(\lambda_0 \phi)}{tanh^2(\lambda_0 \phi)},
\label{phi_derivative_second_pot_SS}
\end{equation}
$C$ denoting a constant of integration.

On solving the above differential equation, the scalar field $\phi(t)$ turns out to be
\begin{equation}
\phi(t) = \frac{1}{4\lambda_0}arcsinh\left(\big[\sqrt{\frac{C_1\rho_c}{6\pi G}}+\frac{C}{4}\sqrt{\frac{V_0}{6\pi G}}\big]\lambda_0(t-t_0)-C_2 \right)
\label{phi_second_pot_SS}
\end{equation}
Here $C_1$ and $C_2$ are integration constants.

The variation of the scalar field with time has been plotted in Figure 11. As we see from the figure, the field $\phi(t)$ has a kink-like shape. Such a shape resembles the spatial variation of a tachyon field used to construct a traversable Lorentzian wormhole\cite{Kar,Sengupta0}, which also incidentally leads to a violation of the NEC to make the wormhole stable by avoiding Weyl curvature singularity at the throat due to infinitely large tidal forces. In this case also, although the NEC is satisfied by the physical matter source in the form of the minimally coupled scalar field, but there is a violation of the NEC by the effective matter arising from contributions due to the quantum geometric corrections.   
\begin{figure}
\centering
\includegraphics[width=0.5\textwidth]{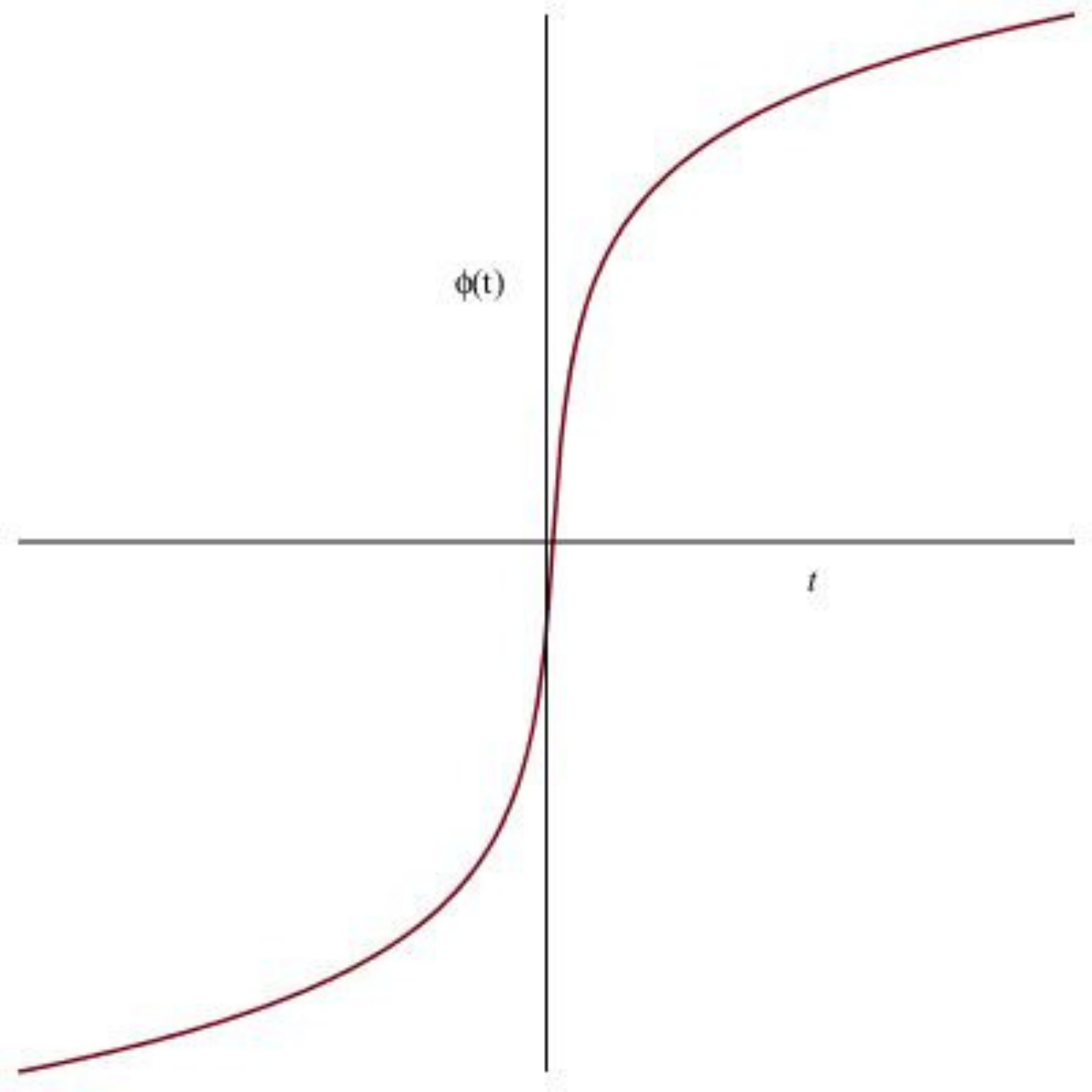}
\caption{Evolution of scalar field $\phi(t)$ with time($t$) for case-3 of LQC.}\label{SS-FIELD2}
\end{figure}

Using the time derivative of the field and the potential, expressing the Hubble parameter in terms of the scale factor and its time derivative, we obtain a differential equation on solving which the scale factor can be expressed in terms of the field as

\begin{equation}
a(\phi)=exp\bigg[-\frac{2\pi G}{C\lambda_0^2}cosh^2(\lambda_0 \phi)+\frac{2\pi GC\prime}{C\lambda_0^2}\bigg]
\end{equation} 

Plugging in the expression for the field and simplyifying, we obtain the scale factor to have the form
\begin{equation}
a(t) = exp\left[\frac{-2\pi G}{C \lambda^2}cosh^2\left\{\frac{1}{4}arcsinh\left(\big[\sqrt{\frac{C_1\rho_c}{6\pi G}}+\frac{C}{4}\sqrt{\frac{V_0}{6\pi G}}\big]\lambda(t-t_0)-C_2 \right)\right\}+\frac{2\pi G C_1}{C \lambda^2}\right]
\label{scale_second_pot_SS_time}
\end{equation}

\begin{figure}
\centering
\includegraphics[width=0.5\textwidth]{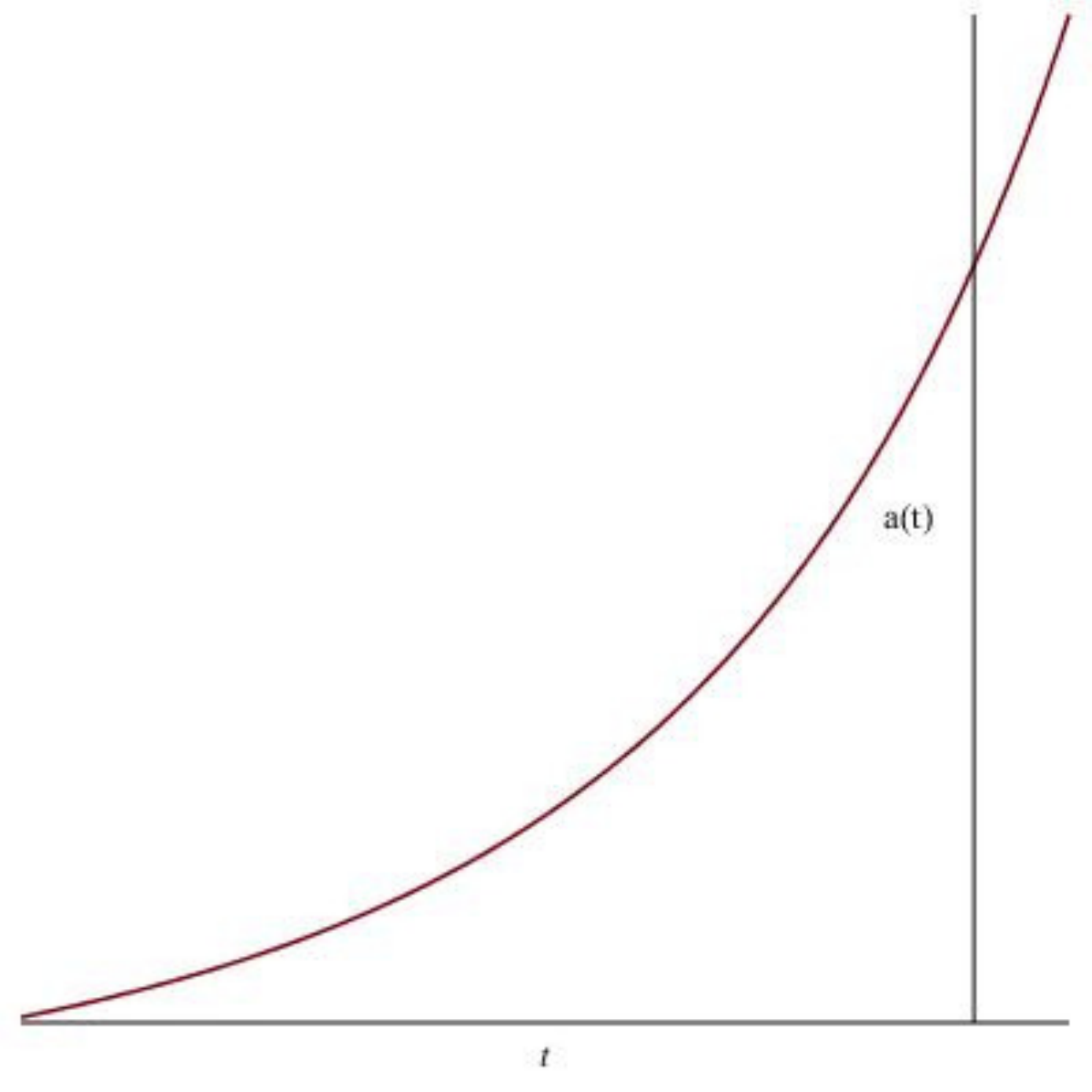}
\caption{Variation of scale factor a(t) with time t for Case-3 of LQC.}\label{SS2}
\end{figure}

The variation of the scale factor with time has been shown in Figure 12. As we can see, the universe is geodesically complete and there is no initial singularity present. The universe emerges into an inflationary state from an eternal quasi-static state and there is a subsequent period of slow-roll inflation. We obtain the slow-roll parameters in the LQC context for the hyperbolic potential and considering sufficient amount of inflation greater than 60 e-folds, we determine the lower limit on the value of the scalar field to begin inflation with.  

The slow roll parameters $\epsilon$ and $\eta$ in LQC turn out to be
\begin{equation}
\epsilon=\frac{1}{2\lambda_p^2}\left(\frac{V^\prime}{V}\right)^2 \frac{1-2\frac{V}{\rho_c}}{\left(1-\frac{V}{\rho_c}\right)^2}
\end{equation}

\begin{equation}
\eta=\frac{1}{\lambda_p^2}\frac{V^{\prime\prime}}{V}\left[\frac{1}{1-\frac{V}{\rho_c}}\right] 
\end{equation}

For our specific choice of the scalar field potential, the slow roll parameters can be computed as

\begin{equation}
\epsilon=\frac{2\rho_c\lambda_0^2sech^4(\lambda_0\phi)}{\lambda_p^2tanh^2(\lambda_0\phi)}\left[\frac{\rho_c - 2V_0tanh^2(\lambda_0\phi)}{\left(\rho_c -V_0tanh^2(\lambda_0\phi)\right)^2}\right]
\label{epsilon_SS}
\end{equation}

\begin{equation}
\eta=\frac{2\lambda_0\rho_c}{\lambda_p^2}\left[\frac{sech^2(\lambda_0\phi)-2tanh^2(\lambda_0\phi)}{sinh^2(\lambda_0\phi)}\right]\left[\frac{1}{\rho_c -V_0tanh^2(\lambda_0\phi)}\right]
\label{eta_SS}
\end{equation}

The number of e-foldings in LQC is given by,

\begin{equation}    
N = \lambda_p^2 \int_{\phi_f}^{\phi_i}\frac{V}{V^\prime}\left(1 - \frac{V}{\rho_c}\right) d\phi
\end{equation}

For our choice of scalar field potential, the number of e-foldings is comuted as

\begin{equation}
N = \frac{\lambda_p^2cosh^2(\lambda_0(\phi_i-\phi_f))}{4\lambda_0^2}\left[1 - \frac{V_0}{\rho_c}\right] + \frac{\lambda_p^2V_0}{2\lambda_0^2\rho_c}ln\left(cosh(\lambda_0(\phi_i-\phi_f))\right)
\label{e-folding_SS}
\end{equation}

The number of e-foldings must be greater than 60 for sufficient inflation to occur. On taking $N>60$, we obtain that $\phi_i>15.178M_5$, where $M_5$ denotes the 5-dimensional Planck mass as before. So $\phi_i$ obtained is considerably lower than in standard GR and in fact lower than the one obtained by us in the braneworld context. So, the inflationary emergent scenario is better realised in LQC than in braneworld or standard GR.

\section{Discussion and Conclusion}

In this paper we have tried to obtain non-singular spatially flat universes in the framework of RS-II braneworld cosmology involving a spacelike extra dimension and in the context of four dimensional loop quantum cosmology in the absence of any effective cosmological constant term. Braneworlds and loop quantum cosmology provide the most efficient frameworks for studying the effective modifications to the standard early universe cosmology obtained within the framework of GR. The initial singularity in cosmology which is often interpreted as the big bang in the standard cosmological model is now thought of by most cosmologists and relativists to be a drawback of GR to explain phenomena involving diverging spacetime curvature and energy densities. This type of a curvature singularity is a Ricci singularity. Had there been diverging tidal forces as in the case of a wormhole throat, it would have been a Weyl curvature singularity. However any curvature singularity is a physical singularity and can not be avoided by a coordinate transformation. This is thought to be undesirable by most physicists and signals the incomptenece of the underlying theoretical framework to be applied to such situations. Due to the very small size of the universe at such early times, quantum effects must be significant and so we may need a quantum gravity (QG) theory to understand such a scenario. However, as we know there is no fully developed and consistent QG theory at this moment, but the two main approaches are the Supertring and M-theories involving extra dimensions and the loop quantum gravity in the usual four spacetime dimensions. For understanding the early universe, we may use effective models inspired from these QG theories, namely the braneworld models and LQC.   

In this paper we are mainly interested about the emergent cosmology models. According to these models, the universe exists eternally in a quasi-static state before evolving into the presently observed state but never reaches a singular state in its process of evolution. Such emergent cosmology models can be constructed in the standard relativistic context with a minimally coupled scalar field at the expense of a postive spatial curvature. We choose two such scalar field potentials which have been applied for this purpose, namely $V(\phi)=A\big(e^{B\phi-1}\big)^2$ and $V(\phi)=V_0 tanh^2(\lambda_0 \phi)$, where $A$, $B$, $\lambda_0$ and $V_0$ are all constant model parameters. As we know, present observations favour a spatially flat universe and so we have attempted to check the viability of emergent cosmological solutions using a minimally coupled scalar field with the above mentioned potentials for a spatially flat universe in the effective frameworks of both RS-II braneworld involving extra dimension and LQC without involving extra dimension. There also exists an emergent cosmology model in the relativistic context for a spatially flat universe, that evolves from a quasi-static state without an inflationary mechanism involving a scalar field but the evolution is described by a perfect fluid with an exotic EoS $p=A\rho-B\rho^\frac{1}{2}$, that can be said to be a mixture of non-baryonic Cold Dark Matter(CDM), exotic matter and dark energy. Here $A$ and $B$ represent the same constant model parameters as those used for the first scalar field potential. We also check whether perfect fluid described by such an exotic EoS can admit a spatially flat emergent cosmology in the braneworld and LQC context also. 

We have considered all the three cases for both the models and obtained the scale factor as solution of the modified first Friedmann equation. The stress-energy conservation equation has been used in the process both for the exotic fluid and the scalar fields. In case of both the scalar field potentials, we have considered weakly decelerated scalar fields which can be characterized by $\frac{dV}{d\phi}<\ddot{\phi}<0$. This immediately makes the $\ddot{\phi}$ term redundant in the equation of motion for the scalar field leading to the simplified result $3H\dot{\phi}+\frac{dV}{d\phi}=0$. This leads to expanding cosmological models corresponding to an EU, immediately leading to a slow-roll inflationary phase in the standard relativistic context.     

We first consider the EoS describing CDM, exotic matter and dark energy. For the RS II brane, the scale factor is found to depend on the EoS parameters $A$ and $B$ and the brane tension $\lambda$. Interestingly, it turns out that on the RS II braneworld the evolution of the Hubble parameter turns out to be singular in the finite past due to vanishing of the scale factor. So, the emergent behaviour can not be obtained on the flat RS II brane. This may be due to the fact that the avoidance of the initial singularity in the relativistic context comes for the violation of the Null energy condition (NEC) by the exotic fluid. However, in the braneworld scenario, what matters is the violation of NEC by the effective matter that arises from the gravitational effects on the brane due to the presence of extra dimension. So, although we take physical matter violating NEC on the brane, but due to the effective matter description NEC is most probalbly restored, resulting in the universe collapsing to a curvature singularity in the finite past. However, in the context of LQC, the obtained scale factor, depending again on the EoS parameters and the critical density $\rho_c$, is non vanishing for all time scales resulting in emergent behaviour arising from a quasi-staic state with finite Hubble parameter. So, an EU for spatially flat universe using the exotic EoS can be realised in LQC but not on the RS II brane.                

In the second case we have investigated the possibility of realizing an emergent cosmology on the brane and LQC with $k=0$, taking a scalar field with exponential potential as the matter source. The obtained scale factors depend on the same factors for both models as in the first case of the exotic EoS. This potential can be said to yield the most interesting result in the sense that for the braneworld we obtain a scale factor representing a bouncing universe rather than an emergent one. The contracting phase is followed by a non-singular bounce leading to a quasi-static state before transiting into an expanding phase. It is worth noting that it is not a matter bounce as we have not used any form of exotic matter that violates the Null Energy Condition (NEC). The bounce can be said to occur due to the extra dimensional gravitational effects leading to the violation of NEC due to effective matter description arising from local quadratic corrections to stress-energy. As the bounce occurs in an universe with zero spatial curvature, so no subsequent inflationary phase is required to bring the universe to its present state. In the context of LQC, we get an even more interesting result in the form of the scale factor describing a cyclic universe. The non-singular bounce mechanism goes on in a repetitive manner. Here also there is no requirement for any inflationary mechanism. For obtaining an accelerating phase, matter must violate the Strong Energy Condition (SEC). However due to quantum effects, a violation of SEC is possible in the context of effective matter description, that allows physical matter source to cause acceleration while conserving the SEC. For a closed universe, in order to avoid an inflationary mechanism, the massive scalar field must play its part in increasing the amplitude of each following cycle, such that the flatness problem can be resolved. However, in our case for a spatially flat universe, the scalar field need not increase the amplitude of each successive cycle.  

Next we have considered the scalar field with a hyperbolic potential as the matter source for both the braneworld and LQC models with vanishing spatial curvature. It turns out that an the scale factor represents an emergent behaviour in both the models for a scalar field with this potential. We go on to obtain the slow roll parameters for both the models. The scale factors, besides depending on the constant parameters involved in the scalar field potential, namely $V_0$ and $\lambda_0$ also depend on the brane tension and the critical density for the RS II braneworld and LQC models, respectively. It is obtained that sufficient inflation with number of e-folds greater than 60 can be obtained with scalar field $\phi_i>34.691 M_5$ for the braneworld model and $\phi_i>15.178 M_5$ for LQC, where $M_5$ represents the 5-dimensional Planck mass. As, $M_5$ is lower than the Planck mass in the usual four dimensions, so the value of $\phi_i$ is also lowered. So, sufficient inflation can be realized with scalar fields of much lower mass to begin with, as compared to standard GR for both models, more so in the case of LQC. This leads to an improved inflationary emergent scenario in both braneworld as well as LQC.

However, we can say that LQC provides a better framework for obtaining non-singular universes than RS-II braneworlds as evident from all three cases. Firstly, for the non-linear EoS, we obtain a non-singular EU scenario from a quasi-static state in the LQC context but in RS II model the universe is geodesically incomplete in the past in finite time. In case of the exponential potential non-singular universe can be obtained in both the cases. In RS-II there is a single non-singular bounce making the universe to transit from a contacting to expanding phase via a quasi-static state. In LQC there are possibly infinite number of non-singular bounces making universe transit from contracting phase to expanding phase and again there is a cosmic turnaround from the expanding to the contracting phase but no intermidiate quasi-static state is present in between. A possible mechanism for this turnaround, as suggested by\cite{Brown} could be that there might be a phantom dark energy, whose energy density keeps on increasing as the universe expands. This makes the term quadratic in energy density also significant at low energies (late-times) which is responsible for the turnaround. In the third and final case where a hyperbolic potential for the scalar field is used, inflationary emergent non-singular universes are obtained in both the contexts with and without extra dimensions, but the mass of the scalar field required to start off the slow roll inflationary mechanism, realizing sufficient inflation with more than 60 e-folds, is lower in LQC than in braneworld. So, a non-singular inflationary emergent model is more favoured in LQC than in braneworld. This can be said due to the presence of the repulsive force in LQC which becomes significant, dominating over the gravitational attraction when spacetime curvature is high enough. There is a violation of NEC by the effective matter arising from the effective quantum correction which is also true in case of a spatially flat universe for correction arising due to a timelike extra dimension which deviates the bulk signature from being Lorentzian. We conclude with the statement that non-singular flat universes are obtained both in the presence of extra dimensions and in standard (3+1)-dimensions accounting for effective quantum corrections.     

\section*{Acknowledgments}

MK and BCP are thankful to the Inter-University Centre for Astronomy and Astrophysics (IUCAA), Pune, India for providing the Visiting Associateship under which a part of this work was carried out. RS is thankful to the Govt. of West Bengal for financial support through SVMCM scheme.

\end{document}